\documentclass[letterpaper, 10 pt, conference]{ieeeconf} 

\usepackage{amsmath}
\usepackage{graphicx}
\usepackage{amssymb}
\usepackage{array}
\usepackage[english]{babel}

\usepackage{caption}
\usepackage{subcaption}
\usepackage{graphicx}
\usepackage{epstopdf}
\usepackage[linesnumbered, boxruled, lined]{algorithm2e}   

\IEEEoverridecommandlockouts                              
\title{\LARGE \bf
On Calculation of Bounds for Greedy Algorithms when Applied to Sensor Selection Problems
}

\author{Jingyuan Liu
}

\begin{document}

\maketitle
\thispagestyle{empty}
\pagestyle{empty}

\begin{abstract}

We consider the problem of studying the performance of greedy algorithm on sensor selection problem for stable linear systems with Kalman Filter. Specifically, the objective is to find the system parameters that affects the performance of greedy algorithms and conditions where greedy algorithm always produces optimal solutions. In this paper, we developed an upper bound for performance ratio of greedy algorithm, which is based on the work of Dr.Zhang \cite{Sundaram} and offers valuable insight into the system parameters that affects the performance of greedy algorithm. We also proposes a set of conditions where greedy algorithm will always produce the optimal solution. We then show in simulations how the system parameters mentioned by the performance ratio bound derived in this work affects the performance of greedy algorithm.

\end{abstract}

\section{INTRODUCTION}

One of the most important problems in dynamics and control is sensor selection problem, which is to select a set of sensors or actuators from a set of options that make the system achieve optimal or near-optimal performance. Since Gauss-Markov linear system model can be used to approximate real world dynamics systems in many cases, an increasing number of works have been done on deriving algorithms for finding optimal or near-optimal solution for sensor selection of such systems and on calculating bounds for performance of those algorithms. In \cite{Dhingra}, Dr. Dhingra successfully applied sparse optimization method on sensor selection problem on continuous time linear dynamical systems and gave an analysis of the algorithm's performance. In \cite{Yang}, Dr. Yang uses convex optimization method to solve the same problem with the additional constraint of uncorrelated sensor noise. In \cite{Sundaram}, Dr. Zhang studied the performance of greedy algorithm when applied to sensor selection problems. 

Zhang summarizes in his paper that greedy algorithm,   which estimates the optimal solution by selecting the sensor or actuator that maximizes benefit for the next step, have comparable performance with the other sensor selection algorithms in general \cite{Sundaram} while having lower complexity and can be applied reliably to a more general class of systems. However, while greedy algorithm works well in most of the cases, its performance tends to drop sharply under certain circumstances. Therefore, in order to utilize greedy algorithm to the best effect, it is important to find out about its limits by inventing methods to evaluate the performance of greedy algorithms.

A number of works have been done on finding ways to evaluate the performance of greedy algorithm. M.Shamaiah's work utilizes the concept of submodularity to produce a performance upper bound for sensor selection problem that have maximizing and submodular cost functions, and provided an example of sensor selection problem with such cost functions. However, since the results in \cite{Shamaiah} have hard constraints in cost function, it can only be applied to a limited range of sensor selection problems.

In \cite{Slavik}, Dr. Slavik improves the existing bound on using greedy algorithm to estimate partial cover and partial d-cover by calculating the bound on each sets in the partial cover independently and then find the the bound on the entire set by summation. In this work, we apply the principle of their method on calculating the performance bound for greedy algorithm for sensor selection problems.

In  \cite{Sundaram}, Dr. Zhang and professor Sundaram proved that it is NP hard to obtain the optimal solution of Kalman filter sensor selection problems, derived a general bound on the performance of all algorithm when applied to sensor selection problems for stable systems, and proved greedy algorithm produces the optimal solution when the sensor noises are uncorrelated and the set of sensor information matrices is totally ordered with respect to positive semi-definiteness. However, since the performance bound derived in Dr.Zhang's paper does not make use of the characteristics of greedy algorithm, it has the risk of being not tight enough when applied to results produced by greedy algorithm, and totally ordered sensor information matrices is a very demanding constraint that many systems can not satisfy.

In this paper, we consider the problem of placing a performance upper bound on greedy algorithm when applied to priori Kalman filtering sensor selection problem (KFSS). Specifically, we study the problem of finding the performance upper bound for greedy algorithm for the systems that can be modeled into stable discrete linear time invariant system models that have uncorrelated sensor noise terms. 

The contribution of this paper is two fold. The first part is to use characteristics of greedy algorithm and the current bound to construct a tighter upper bound for greedy performance when applied to priori KFSS problems. For priori KFSS problems, we show that the performance of sensors selected by greedy algorithm can be bounded by a parameter that depends only on the system dynamics matrix, the state measurement matrix, and the noise covariance matrices for both measurement and state noise. 

Our second contribution is to propose a condition for the system model which allows the sensor information matrices $R_i$ to be totally ordered. Since Zhang proved in his work \cite{Sundaram} that if $R_i$ are totally ordered and the sensor noises are uncorrelated then greedy algorithm produces the optimal result, our contribution proves that if the parameters of a system satisfies certain constraint then {\it a priori} covariance based greedy algorithm can produce the optimal result for the system.

The rest of the paper is organized as follows. A formulation of the sensor selection problem is presented in section II. Section III analyzes Zhang's method of calculating bound for sensor selection problem. Section IV proposes the performance upper bound for greedy algorithm when applied to priori KFSS problem, and section VI summarize the paper and offer suggestions on future work.

\section{Problem Formulation}

Consider the discrete time linear system model

\begin{equation}
 \label{e:dl}
 x_{k+1}=Ax_k+\omega_k
\end{equation}

Where $A$ is the system dynamics matrix and $\omega_k$ is the state noise associated with step k. For sensor selection, $q$ sensors are to be chosen from a given set of sensors $Q$, where $|Q| \geq q$. For individual sensor $i$, the state measurement matrix is denoted as $C_i$, and the corresponding noise is $v_i$, which for this work is assumed to be zero mean white Gaussian noise process. Then for sensor $i$ we have:

\begin{equation}
 \label{e:dlsi}
 y_{k}^i=C_i x_k+v_k^i
\end{equation}

If we define $y_k$ as the collective output of all the sensors for step $k$, $C$ as the state measurement matrix of all the sensors, and $v_k$ as the vector of all the perturbations on sensor at step $k$, we have:

\begin{equation}
 \label{e:dls}
 y_{k}=C x_k+v_k
\end{equation}

We assume the pair $(A,C)$ is detectable, and define $V$ as the error covariance matrix for sensor noise of all sensors in $Q$. Since the system model used here is linear and no non-Gaussian noise exists in this model, we use {\it a priori} steady state Kalman filter for state estimation. We define $z \in \{ 0,1 \}^q$ as the indicator vector of the selected sensors. That is, $z_i=1$ if and only if sensor $i \in Q$ is being selected. Then, by using Dr. Zhang's definition of selection matrix $Z \triangleq diag(z_1 I_{s_1 \times s_1},...z_q I_{s_q \times s_q})$ \cite{Sundaram}, we define $\tilde{C_z} \triangleq ZC$ as the state measurement matrix for chosen sensors, and $\tilde{V_z} \triangleq ZVZ^T$ as the error covariance matrix for sensor noise of chosen sensors. For evaluating the performance of the set of chosen sensors, we use information from steady state Discrete Algebratic Riccati Equation (DARE) \cite{Anderson}:

\begin{equation}
 \label{e:dare}
 \Sigma(z)=W+A(\Sigma^{-1}(z)+\tilde{C_z}^T \tilde{V_z}^{-1} \tilde{C_z})^{-1} A^T
\end{equation}

Where $\Sigma(z)$ is the error covariance matrix of steady state Kalman filter for sensor selection vector $z$. and $W$ is the error covariance matrix for state noise. We also assume that the pair $(A, W^\frac{1}{2})$ is stabilizable.

For choosing the sensor selection that yields optimal results, the cost function $J(z)$ is designed to measure the effectiveness of selection.

\begin{equation}
 \label{e:cost}
 J(z)=trace(\Sigma(z))
\end{equation}

For finding the sub-optimal approximation of the set of sensors that yields the smallest cost, this work uses the same {\it a priori} covariance based greedy algorithm used by professor Sundaram in his work, the process of which is detailed below \cite{Sundaram}.

\begin{algorithm}
\caption{{\it A Priori} Covariance based Greedy Algorithm } \label{Algorithm_1} 

Input: System dynamics matrix A, set of all sensors Q, noise covariance matrices W and V, and number of sensors to be chosen q

Output: A set S of chosen sensors

$k\leftarrow 0$, $S\leftarrow \emptyset$

for $k\leq q$ do

for $i\in Q\cap \overline{S} $ do

Calculate $J(i,S)=J(S\cup {i})$

end for

Choose j that, depends on the requirement, maximizes or minimizes $J(i,S)$

$S \leftarrow {j},Q \leftarrow Q \backslash {j},k\leftarrow k+1$

end for

\end{algorithm}

\vspace{5mm}

Defining $z_{selected}$ as the set of sensors chosen by sensor selection algorithm and $z_{opt}$ as the set of sensors that leads to optimal performance. Then we introduce the concept of performance ratio $r_{greedy}(\Sigma)$, which is defined by

\begin{equation}
 \label{e:costratio}
 r(\Sigma)=\frac{J(z_{selected})}{J(z_{opt})}=\frac{trace(\Sigma(z_{selected}))}{trace_(\Sigma(z_{opt}))}
\end{equation}

to evaluate the performance for sensor selection algorithms.

For determining the feasibility of the sensor selection, we use Anderson's result on Kalman Filtering \cite{Anderson}:

\vspace{3mm}

\textbf{Lemma 1}     
When the pair $(A,\tilde{C_z})$ is detectable and the pair $(A,W^{\frac{1}{2}})$ is stabilizable, then the result of sensor selection algorithm is feasible.

\vspace{3mm}

Now we can describe the problem of finding a metric to evaluate the performance of greedy algorithm when applied on sensor selection problems with {\it a priori} Kalman Filter. Given a system dynamics matrix $A \in R^{n \times n}$, a measurement matrix $C \in R^{|Q| \times n}$, a positive semi-definite system noise covariance matrix $W \in R^{n \times n}$, and a positive semi-definite sensor noise covariance matrix $V \in R^{n \times n}$, we then use greedy algorithm to find the sub-optimal approximation to the solution to the following problem:

\vspace{3mm}
$\min\limits_{z} J(z)$\\
\vspace{2mm}
$z \in \{ 0,1 \}^q$\\
\vspace{3mm}

Where $J(z)$ is given by Eq. \ref{e:cost}, or else determine that no feasible sensor selection exists. 

Defining the indicator vector of the sub-optimal sensor set selected by greedy algorithm as $z_{greedy}$ and the indicator vector of the solution of this problem as $z_{opt}$, our problem is to study how system parameters affects the performance ratio of greedy algorithm $r_{greedy}(\Sigma)=\frac{J(z_{greedy})}{J(z_{opt})}$.

\section{Bound on Greedy Performance}
In this section, we discuss some of existing methods of calculating performance guarantees of greedy algorithm as functions of system parameters. 

\subsection{Submodularity}

One such tool is the  concept of submodularity, which was first applied to the problem of evaluating the performance of greedy algorithm on sensor selection problem by M.Shamaiah, as mentioned in the introduction. To be more specific, Dr. Shamaiah proved that for sensor selection problems with submodular and maximizing cost functions, the greedy algorithm always arrive at a solution which cost is within a factor of $1-\frac{1}{e}$ of the optimal solution. We can refer to \cite{Shamaiah} for a detailed discussion of the proofs and examples.

The concept of submodularity is defined as follows:

\vspace{5mm}
\textbf{Definition 1} [submodularity]

\textit{Let Q be any non-empty finite set, and $2^{Q}$ be its power set. Then a set function $f:2^{Q}\rightarrow R^{+}$ is submodular if and only if:
}

\textit{
$\forall X_{1},X_{2} \subseteq Q$, if $X_{1} \subseteq X_{2}$ then for every $x\in X_{1} \setminus X_{2}$
},

\textit{
$f(X_{1} \cup \{ x \})-f(X_{1})-f(X_{2} \cup \{ x \})-f(X_{2}) \geq 0$
}
\vspace{5mm}

That is, the function f has diminishing increments property. 

While \cite{Shamaiah} offers an example where the cost function is submodular for all conditions, and \cite{Jawaid} offers several examples of cost functions that are submodular in some circumstances, \cite{Sundaram} shows that the cost functions proposed in \cite{Jawaid} are not submodular in general. This reinforces the point that submodular cost function is a very tight constraint that a lot of systems would not be able to satisfy.

\subsection{Zhang's method}

Another tool of this kind is Zhang's method \cite{Sundaram}, which provides an estimation to the performance of all sensor selection algorithms when the system dynamics matrix is stable. This method constructs an upper bound to the performance ratio $r_{greedy}(\Sigma)$ by finding the upper bound of the cost of solution of greedy algorithm $trace(\Sigma_{greedy})$ and the lower bound of the cost of optimal solution $trace(\Sigma(z_{opt}))$, then dividing the upper bound by the lower bound.
z
To find the upper bound to the cost of the result of all algorithms, Zhang utilizes the principle that no set of selected sensor will have a higher cost than the case where no sensor is selected. Since the system dynamics matrix is stable, DARE has solutions even when no sensor is selected. The no sensor solution of DARE, $\Sigma(0)$, can be calculated from the following equation.

\begin{equation}
 \label{e:dareu}
 \Sigma(0)=A\Sigma(0)A^T+W
\end{equation}

From this equation, the upper bound for $trace(\Sigma(z_{selected}))$ can be derived as:

\begin{equation}
 \label{e:upper}
 trace(\Sigma(z_{selected}))\leq trace(\Sigma(0)) \leq \frac{\sigma_1^2 (P)}{\sigma_n^2 (P)} \frac{trace(W)}{1-\sigma_1^2 (D)}
\end{equation}

Where $\sigma_1$ is the largest singular value of the matrix, $\sigma_n$ is the smallest singular value of the matrix, $P$ is a nonsingular matrix that satisfies $\sigma_1 (PAP^{-1}) \leq 1$, and $D \triangleq (PAP^{-1})$.

We denote the set of first k sensors selected by sensor selection algorithm as $z_k$, where $k \in [0,q]$. The lower bound for the denominator in the performance ratio equation, $trace(\Sigma(z_{opt}))$, is then derived from the following inequality which can be satisfied by any given $z_k$, where $R_k$ is the sensor information matrix corresponding to $z_k$. 

\begin{equation}
 \label{e:lowerdenom1}
 trace(\Sigma(z_k))\geq trace(A(W^{-1}+R_k)^{-1}A^T +W)
\end{equation}

\begin{equation}
 \label{e:lowerdenom2}
 trace(\Sigma(z_k))\geq \lambda_{n}(A^T A)trace(W^{-1}+R_k^{-1})+trace(W)
\end{equation}

\begin{equation}
 \label{e:lowerdenom3}
 trace(\Sigma(z_{opt}))\geq \frac{n \sigma_n^2 (A)}{\lambda_1 (W^{-1}+R_k)}+trace(W)
\end{equation}

\begin{equation}
 \label{e:lower}
 trace(\Sigma(z_{opt}))\geq \frac{n \sigma_n^2 (A)}{\lambda_1 (W^{-1})+\lambda_1 (R_k)}+trace(W)
\end{equation}

Where $\lambda_1$ is the largest eigenvalue of the matrix, and $R_k$, the sensor information matrix corresponding to the indicator vector $z_k$, is defined as $\tilde{C_k}^T \tilde{V_k}^{-1} \tilde{C_k}$.

The upper bound for the performance ratio is then obtained by dividing the upper bound of $trace(\Sigma(z_{selected}))$ by the lower bound of $trace(\Sigma(z_{opt}))$.

\subsection{Zhang's proposition}

One other method is to evaluate performance of greedy algorithm by using the relationship between the sensor information matrices $R_{i}$ and \textit{priori} Kalman Filter error covariance matrices. In \cite{Sundaram}, Dr. Zhang proposed that for both \textit{priori} and \textit{a priori} KFSS problems, the solution provided by greedy algorithm would be the optimal solution if the sensor noises are uncorrelated and $\{ R_{i} \}$ are totally ordered with respect to the order relation of positive semi-definiteness. 

\section{Our Contribution}

In this section, we propose a method of deriving performance bounds based on the the method proposed by H. Zhang, and a category of systems where Zhang's proposition is applicable. Compare to the Zhang's method,our method of calculating performance ratio upper bound offers a tighter bound by improving on the upper bound for the cost of the sensor set selected by the greedy algorithm. As proven by the simulation results in section V, the parameters highlighted by the performance bound affects the performance of the greedy algorithm greatly. In the rest of the section, we propose a condition where the sensor information matrices $R_i$ are totally ordered. Then from Zhang's proposition, we can conclude that greedy algorithm produces optimal solutions when applied to the systems that satisfy those conditions.

\subsection{Proposed Method for Performance Ratio Upper Bound Calculation}

The method we proposed is based on the following fundamental characteristics of greedy algorithm:

\begin{itemize}
\item The cost is reduced each time a new sensor is chosen.
\item For each iteration, greedy algorithm chooses the sensor that leads to maximum cost reduction for the next step.
\end{itemize}

It can be applied to the systems that satisfies the following assumptions:

\vspace{3mm}
\textbf{Assumption 1}

The Kalman error covariance matrix for each iteration of greedy algorithm is totally ordered with respect to the order relation of positive semi-definiteness. That is, if we define $z_k$ as the set of sensors selected by the first $k \in [0,q]$ iteration of greedy algorithm, then $\Sigma(z_k) \geq \Sigma(z_{k+1})$.

\vspace{3mm}

\textbf{Assumption 2}

The sensor noise is uncorrelated. That is, the sensor noise covariance matrix is block diagonal.

\vspace{3mm}

We define the change of cost when the $(k+1)^{th}$ sensor is selected as $\triangle J(z_{k+1})$, which is always negative due to the nature of the cost function we selected. We also define $J(z_k)$ as the cost with $k^{th}$ sensor is chosen by greedy algorithm. Then we can write $\triangle J(z_{k+1})$ as:
\vspace{3mm}

\begin{equation}
 \label{e:Jdelta1}
 \triangle J(z_{k+1})=J(z_{k+1})-J(z_{k})
\end{equation}

If we define $J(z_1)$ as the cost when the first sensor is selected and $J(z_{greedy})$ as the cost of sensors selected by greedy algorithm, $J(z_{greedy})$ can then be expressed as follows,

\begin{equation}
 \label{e:greedycost}
 J(z_{greedy})=J(z_1)+\sum_{k=1}^{q-1} \triangle J(z_{k+1})
\end{equation}

Therefore, if we can find a matrix $M$ such that $trace(M) \geq J(z_1)$ for all cases and a value $\alpha$ such that $\alpha \geq \triangle J(z_{k+1})$ for all cases, then we have $J(z_{greedy}) \leq trace(M)+(q-1) \alpha$.

For the matrix M, If the systems are stable and we prioritize saving computational resources over the tightness of bound calculated, we can simply use $trace(\Sigma(0))$ for $trace(M)$, since by the fundamental characteristics of greedy algorithm, under no condition the cost can be higher than the case when no sensor is selected. However, if we can afford to improve bound tightness at the expense of computation resources or we need to find the bound of performance for unstable systems, we can simply find all the possible cases for $J(z_1)$ and choose the worst case manually. In both cases, we have a matrix $M$ that satisfies the condition $trace(M) \geq J(z_1)$. Which leads to the following inequality,

\begin{equation}
 \label{e:greedycost1}
 J(z_{greedy}) \leq trace(M)+\sum_{k=1}^{q-1} \triangle J(z_{k+1})
\end{equation}

As for the derivation of $\alpha$, first step is to write out the standard expression for $\triangle J(z_{k+1})$. From Equation \ref{e:dare} and \ref{e:cost} we can obtain the expression for $J(z_{k+1})$ as:

\begin{equation}
\begin{split}
 \label{e:Jkz1}
 J(z_{k+1})=trace(A[\Sigma(z_{k+1})^{-1}+\tilde{C}_{z_{k+1}}^T \tilde{V}_{z_{k+1}}^{-1} \tilde{C}_{z_{k+1}}]^{-1}\\ A^T)
 +trace(W)\\
 =trace(A[\Sigma(z_{k+1})^{-1}+\tilde{C}_{z_{k}}^T \tilde{V}_{z_{k}}^{-1} \tilde{C}_{z_{k}}+\triangle_{k}]^{-1}\\ 
 A^T)+trace(W)
\end{split}
\end{equation}

Where $\tilde{C_{z_k}}$ and $\tilde{V_{z_k}}$ are the state measurement matrix and sensor noise covariance matrix for up to $k^{th}$ sensors selected by greedy algorithm respectively. Also, $\triangle_{k}=\tilde{C}_{z_{k+1}}^T \tilde{V}_{z_{k+1}}^{-1} \tilde{C}_{z_{k+1}}-\tilde{C}_{z_{k}}^T \tilde{V}_{z_{k}}^{-1} \tilde{C}_{z_{k}}$.

And 

\begin{equation}
\begin{split}
 \label{e:Jkz}
 J(z_{k})=trace(A[\Sigma(z_{k})^{-1}+\tilde{C}_{z_{k}}^T \tilde{V}_{z_{k}}^{-1} \tilde{C}_{z_{k}}]^{-1} A^T)+\\
 trace(W)
\end{split}
\end{equation}

Then we have,

\begin{equation}
\begin{split}
 \label{e:Jdelta2}
 \triangle J(z_{k+1})=trace(A \Big[ [\Sigma(z_{k+1})^{-1}+\tilde{C}_{z_{k}}^T \tilde{V}_{z_{k}}^{-1} \tilde{C}_{z_{k}}+\triangle_{k}]^{-1}-\\
 [\Sigma(z_{k})^{-1}+\tilde{C}_{z_{k}}^T \tilde{V}_{z_{k}}^{-1} \tilde{C}_{z_{k}}]^{-1} \Big] A^T)
\end{split}
\end{equation}

By using the Matrix Inequality Lemma proven in appendix F, we can arrive at the following inequality from assumption 1:

\begin{equation}
\begin{split}
\label{e:Jdelta3}
\triangle J(z_{k+1}) \leq trace(A \Big[ [\Sigma(z_{k})^{-1}+\tilde{C}_{z_{k}}^T \tilde{V}_{z_{k}}^{-1} \tilde{C}_{z_{k}}+\triangle_{k}]^{-1}-\\
 [\Sigma(z_{k})^{-1}+\tilde{C}_{z_{k}}^T \tilde{V}_{z_{k}}^{-1} \tilde{C}_{z_{k}}]^{-1} \Big] A^T)
\end{split}
\end{equation}

Recalling the definition of sensor information matrix in the Section III, we can write $R_{k+1}=\tilde{C}_{z_{k+1}}^T \tilde{V}_{z_{k+1}}^{-1} \tilde{C}_{z_{k+1}}$. To facilitate derivation, we define $B_{k+1}=R_{k+1}-R_k$ as the change in sensor information matrix when the $k+1^{th}$ sensor is selected. This leads us to the following equation,

\begin{equation}
\label{e:Rk}
R_k=\tilde{C}_{z_{k}}^T \tilde{V}_{z_{k}}^{-1} \tilde{C}_{z_{k}}=\sum_{i=1}^k B_i
\end{equation}

\begin{equation}
\label{e:Rk1}
R_{k+1}=\tilde{C}_{z_{k+1}}^T \tilde{V}_{z_{k+1}}^{-1} \tilde{C}_{z_{k+1}}=\sum_{i=1}^k B_i+B_{k+1}
\end{equation}

Then, if we define $D_{ij}=\Sigma(z_{i})^{-1}+\sum_{m=1}^j B_m$ and apply Miller's Theorem of Matrix Inversion \cite{Miller} on $[\Sigma(z_{k})^{-1}+\sum_{i=1}^k B_i +B_{k+1}]^{-1}-[\Sigma(z_{k})^{-1}+\sum_{i=1}^k B_i]^{-1}$ we have,

\begin{equation}
\begin{split}
\label{e:res1}
-\frac{[D_{k,k}^{-1} B_{k+1} D_{k,k}^{-1}]}{1+trace(D_{k,k}^{-1} B_{k+1})}=[\Sigma(z_{k})^{-1}+\sum_{i=1}^k B_i +B_{k+1}]^{-1}-\\
 [\Sigma(z_{k})^{-1}+\sum_{i=1}^k B_i]^{-1}
\end{split}
\end{equation}

Since $1+trace(D_{k,k}^{-1} B_{k+1})$ is a real number, if we can find real numbers $u$ and $c$ such that $c \geq trace(D_{k,k}^{-1} B_{k+1})$ and $trace(A[D_{k,k}^{-1} B_{k+1} D_{k,k}^{-1}] A^T)-u \geq 0$ for all $D_{k,k}$ and $B_{k+1}$, we can write,

\begin{equation}
\label{e:res2}
trace(\frac{A[D_{k,k}^{-1} B_{k+1} D_{k,k}^{-1}]A^T}{1+trace(D_{k,k}^{-1} B_{k+1})}) \geq \frac{u}{1+c}
\end{equation}

From Equations \ref{e:Jdelta3} and \ref{e:res1}, we have,

\begin{equation}
\label{e:res3}
\triangle J(z_{k+1}) \leq -trace(A \frac{[D_{k,k}^{-1} B_{k+1} D_{k,k}^{-1}]}{1+trace(D_{k,k}^{-1} B_{k+1})} A^T) \leq -\frac{u}{1+c}
\end{equation}

To find a suitable candidate of $c$, we need to go back to DARE, 

\begin{equation}
 \label{e:dare1}
 \Sigma(z_k)=A D_{k,k}^{-1} A^T+W
\end{equation}

Since $M-\Sigma(z_k)$ is positive semi-definite for all possible $z_k$, From the Matrix Inequality Lemma 1 in Appendix B we have,

\begin{equation}
 \label{e:dare2}
 A^{-1} (M-W) A^{-T}A \succeq D_{k,k}^{-1}
\end{equation}

Here $\succeq$ represents if the matrix on the left side is subtracted from the matrix on the right side the resulting matrix will be positive semi-definite.

Since we can always find a set of sensor $z_K$ such that $B_{z_K} \succeq B_{k+1}$ for all possible $B_{k+1}$, by using Matrix Inequality Lemma 2 we can write the inequality,

\begin{equation}
 \label{e:dare3}
 A^{-1} (M-W) A^{-T}A B_{z_K} \succeq D_{k,k}^{-1} B_{z_K}
\end{equation}

Which gives us the acceptable c value $trace(A^{-1} (M-W) A^{-T}A B_{z_K})$. 

As for an acceptable expression for $u$, we can rely on the property mentioned in
Dr. Fang's work\cite{Fang}.

\begin{equation}
\begin{split}
\label{dare35}
\lambda_{min} (D_{k,k}^{-1} A^T A D_{k,k}^{-1} ) trace(B_{k+1}) \leq \\ 
trace(A[D_{k,k}^{-1} B_{k+1} D_{k,k}^{-1}] A^T)
\end{split}
\end{equation}

With the Matrix Inequality Lemma 3, which is proved in the appendix, we can write,

\begin{equation}
\label{dare37}
(\lambda_{min} (A) \lambda_{min} (D_{k,k}^{-1}))^2 \leq\lambda_{min} (A D_{k,k}^{-1})^2\leq \lambda_{min} (D_{k,k}^{-1} A^T A D_{k,k}^{-1} )
\end{equation}

For an lower bound on $\lambda_{min} D_{k,k}^{-1}$, we can use the property that  $\Sigma(z_k) \succeq W$ for any $k$ in all selected set of sensors $z$. Therefore, by finding a matrix $B_{\overline{z_R}}$ such that $B_{\overline{z_R}} \succeq \sum_{i=1}^{p-1} B_i \succeq \sum_{i=1}^{k-1} B_i$ for all possible $z$ for the given $p$, we can have 

\begin{equation}
 \label{e:dare4}
 D_{k,k}=\Sigma(z_k)^{-1}+\sum_{i=1}^{k} B_i \preceq W^{-1}+B_{\overline{z_R}}
\end{equation}

Which leads to $D_{k,k}^{-1} \succeq [W^{-1}+B_{\overline{z_R}}]^{-1}$. Then, according to Matrix Inequality Lemma 4, which is proved in the appendix, $\lambda_{min} (D_{k,k}^{-1}) \geq \lambda_{min} ([W^{-1}+B_{\overline{z_R}}]^{-1})\geq 0$.

If we can find a real number b such that $b \leq trace(B_{k+1})$ for all possible $B_{k+1}$, we can have an acceptable expression for u,

\begin{equation}
 \label{e:dare5}
 u=(\lambda_{min} (A) \lambda_{min} ([W^{-1}+B_{\overline{z_R}}]^{-1}))^2 b
\end{equation}

From Equations \ref{e:greedycost1} and \ref{e:res3} we can arrive at the following upper bound for the result of greedy algorithm:

\begin{equation}
\begin{split}
 \label{e:ubound}
  \triangle J(z_{k+1})=trace(M)+\sum_{k=2}^{q} \triangle J(z_{k+1}) \leq \\
 trace(M)-(q-1)*\frac{u}{1+c}
 \end{split}
\end{equation}

Which, combined with the lower bound for the cost of optimal result from Equation \ref{e:lower}, leads to an improved upper bound on the performance ratio of greedy algorithm results,

\begin{equation}
\label{e:ratiobound}
r_{greedy}(\Sigma) \leq r_{new}(\Sigma) \triangleq \frac{trace(M)-(q-1)*\frac{u}{1+c}}{\frac{n \sigma_n^2 (A)}{\lambda_1 (W^{-1})+\lambda_1 (R_k)}+trace(W)}
\end{equation}

Alternatively, it can also be written as the lower bound of performance improvement from the greedy algorithm, which is a self-contained function of system parameters, subtracted from the performance bound derived by Dr.Zhang.

Since the performance bound derived by Dr.Zhang is \cite{Sundaram},

\begin{equation}
\label{e:oldbound}
r_{old}(\Sigma) = \frac{\alpha_A (1+\lambda_1^{max} \lambda_n(W)) trace(W)}{n \sigma_n^2 (A) \lambda_n (W)+(1+\lambda_1^{max} \lambda_n(W)) trace(W)}
\end{equation}

From Equations \ref{e:dare3}, \ref{e:dare5} and \ref{e:oldbound}, we arrive at this theorem,
\vspace{3mm}

\textbf{Theorem 1}

For any cost vector $ r $ and maximum allowed cost $ \beta $, we design $ \textbf{R} =\{ R_k \} $ as the set of sensor information matrix that satisfies the constraint $ r^T z \leq \beta $. Then, we design its subsets  \textbf {$R_1$}  and \textbf {$R_q$}  as set of individual sensor information matrix and set of sensor information matrix for q sensors, respectively. Denote $\lambda_1^{max} \triangleq max\{ \lambda_1(R) \| R \in \textbf{R} \}$ and $b \triangleq min\{ trace(R) \| R \in $ \textbf{$R_1$} $ \}$. Then for any stable systems with positive definite matrix $W$,

\vspace{3mm}

\begin{equation}
\begin{split}
\label{e:newbound}
r_{greedy}(\Sigma) \leq r_{new}(\Sigma)=r_{old}(\Sigma)- (q-1)*\\
\frac{ \sigma_n^2 (A) \sigma_n ^2 ((W^{-1} +B_{\overline{z_R}})^{-1}) (\lambda_1 (W^{-1})+\lambda_1^{max}) b}{(1+trace(\Sigma (0) B_{z_K}))(n \sigma_n^2 (A)+trace(W)(\lambda_1 (W^{-1})+\lambda_1^{max}))} 
\end{split}
\end{equation}
 
Where $B_{\overline{z_R}}$ can be any matrix that satisfies $B_{\overline{z_R}}-R \succeq 0$ for all $R \in$ \textbf {$R_q$}, and $B_{z_K}$ can be any matrix that satisfies $B_{z_K}-R \succeq 0$ for all $R \in$ \textbf {$R_1$}.

This result can also lead to a simpler upper bound for $r_{greedy}(\Sigma)$ which highlights the effect of system dynamics matrix $A$, number of states $n$ and the sensor information matrix for all the available sensors $R_{all} \triangleq C^T V^{-1} C$ on the performance of greedy algorithm. 

\vspace{3mm}

\textbf{Corollary 1}

For any stable system with positive definite W, 

\begin{equation}
\begin{split}
\label{e:newboundsimp}
r_{greedy}(\Sigma) \leq r_{old}(\Sigma)- (q-1)*\frac{1}{1+ \| R_{all} \|_2 trace(\Sigma (0))}*\\
\frac{1}{(\|W^{-1} \|_2 +\| R_{all} \|_2)^2}*\frac{b}{\frac{n}{\| (W^{-1})\|_2+\lambda_1^{max}}+\frac{trace(W)}{\sigma_n^2 (A)}}
\end{split}
\end{equation}

Where $\| R_{all} \|_2 \triangleq \sqrt{\lambda_1 (R_{all}^T R_{all})}$ denotes the spectral norm of the matrix $R_{all}$ 
\vspace{3mm}

\textbf{Proof}

Since the sensor noise is uncorrelated, $R_{all}$ satisfies $R_{all}-R \succeq 0$ for all $R \in$ \textbf {$R_q$} and $R \in$ \textbf {$R_1$}. Therefore, $R_{all}$ is a suitable candidate for both $B_{\overline{z_R}}$ and $B_{z_K}$. Replacing both matrices by $R_{all}$ leads to,

\begin{equation}
\begin{split}
\label{e:newboundsimp1}
r_{greedy}(\Sigma) \leq r_{old}(\Sigma)- (q-1)*\frac{1}{1+ trace(\Sigma (0) R_{all})}*\\
\frac{1}{(\sigma_1 ^2 (W^{-1} +R_{all}))^2}*\frac{b}{\frac{n}{\lambda_1 (W^{-1})+\lambda_1^{max}}+\frac{trace(W)}{\sigma_n^2 (A)}}
\end{split}
 \end{equation}

From the definition of norm, the spectral norm of a matrix $\| . \|_2$ satisfies the triangle inequality principle. Therefore, $\| W^{-1}+R_{all} \|_2 \leq \| W^{-1} \|_2 +\| R_{all} \|_2$, and since the spectral norm of a matrix is always positive, $\frac{1}{\sigma_1 ^2 (W^{-1} +R_{all})}=\frac{1}{(\| W^{-1}+R_{all} \|_2)^2} \geq \frac{1} {(\| W^{-1} \|_2 +\| R_{all} \|_2)^2}$. Also, from Dr. Patel's work \cite{Patel} we have $\frac{1}{trace(\Sigma (0) R_{all})} \geq \frac{1}{\| R_{all} \|_2 trace(\Sigma(0))}$. Applying these inequalities to Equation \ref{e:newboundsimp1} gives us Corollary 1.
\vspace{3mm}

\textbf{Remark 1}

From Corollary 1 we can observe that if every other elements in Equation \ref{e:newboundsimp} remains the same, the lower bound on greedy algorithm's performance ratio reduction is higher for systems with lower $\| R_{all} \|_2$ or smaller dimension $n$. Since the systems discussed in this paper have uncorrelated sensor noise, the lower and upper bound of $\| R_{all} \|_2$ decreases when $ |Q| $ decreases. Therefore, for stable systems with uncorrelated sensor noise, greedy algorithm performs better when $|Q|$ decreases  .

\vspace{3mm}

\subsection{On Zhang's Proposition}
Our work on Zhang's proposition centers around finding conditions of system parameter for KFSS problems where greedy algorithm always produces the optimal solution.  
\vspace{3mm}

\textbf{Lemma 2}

Greedy Algorithm always produces the optimal solution for KFSS problems where the sensor noise are uncorrelated, the column spaces of $C_i^T C_i$ are the same for all $i \in Q$ and for all $i \geq j$ $\frac{S_i^2}{V_{i,i}} \geq \frac{S_j^2}{v_{j,j}}$, where $S_i$ is the only non zero singular value of $C_i$. 
\vspace{3mm}

\textbf{Proof}

When the sensor noises are uncorrelated, one condition that allows $\{ R_i \}$ to be totally ordered is for all $C_i^T V_{i,i}^{-1} C_i-C_j^T V_{j,j}^{-1} C_j$ to be positive semi-definite for all $i \geq j, i,j \leq |Q|$. That is, if for all vector x in $R^n$, $x^T (C_i^T V_{i,i}^{-1} C_i-C_j^T V_{j,j}^{-1} C_j)x \geq 0$ then $\{ R_i \}$ are totally ordered.

Since $C_i$ are n by 1 vectors, $C_i^T C_i$ will be rank 1 n by n matrices, which means if the column spaces of $C_i^T C_i$ are the same then their null spaces must be the same, and all $C_i^T C_i$ have the same eigenvectors. Since $V_{i,i}$ are scalars, if the above condition is satisfied then all $R_i$ have the same eigenvectors. 

We also notice that since $C_i^T V_{i,i}^{-1} C_i$ are n by n square matrices, every possible $x$ can be written as a combination of eigenvectors that corresponds to its both zero and non zero eigenvalues. Also, since $C_i^T V_{i,i}^{-1} C_i$ are symmetric their column spaces are the same as their row spaces and their left null spaces are the same as their right null spaces, and that because $S_i^2$ is the only non zero eigenvalue of $C_i^T C_i$, the expression of the only non zero eigenvalue of $C_i^T V_{i,i}^{-1} C_i$ is $\frac{S_i^2}{V_{i,i}}$.

Based on those derivations, we now can say if $w^T (\frac{S_i^2}{V_{i,i}}-\frac{S_j^2}{V_{j,j}})w \geq 0$ then $\{ R_i \}$ are totally ordered, where $w$ is the only eigenvector that correspond to non-zero eigenvalue for both column space and row space. Since $w^T w$ is always larger than 0, if $\frac{S_i^2}{V_{i,i}}-\frac{S_j^2}{V_{j,j}} \geq 0$ then $ \{ R_i \}$ are totally ordered, and from Zhang's proposition \cite{Sundaram} we can conclude that under these circumstances greedy algorithm yields the optimal solution.

\section{Simulation}
In this section, we compare the simulation results for the performance of {\it a priori} Kalman filter error covariance matrix based greedy algorithm when applied to sensor selection problems for systems with different parameters.

For this simulation, we randomly generate 600 systems that have the following characteristics each run,

\begin{itemize}
\item A total number of 4 sensors are selected for each system. ($q=4$)
\item All systems are stable, and all of its matrices have real valued elements. However, some systems have complex conjugate eigenvalue pairs.
\item All systems have uncorrelated sensor noise. That is, the sensor noise covariance matrices for all systems are block diagonal.
\end{itemize}

In order to demonstrate the effects of system parameters discussed in this paper has on the performance in greedy algorithm, we run greedy algorithm on following categories of systems,

\begin{itemize}
\item $n=5, |Q|=10$
\item $n=5, |Q|=20$
\item $n=5, |Q|=30$
\end{itemize}

\begin{figure}[htb]
 \includegraphics[width=90mm]{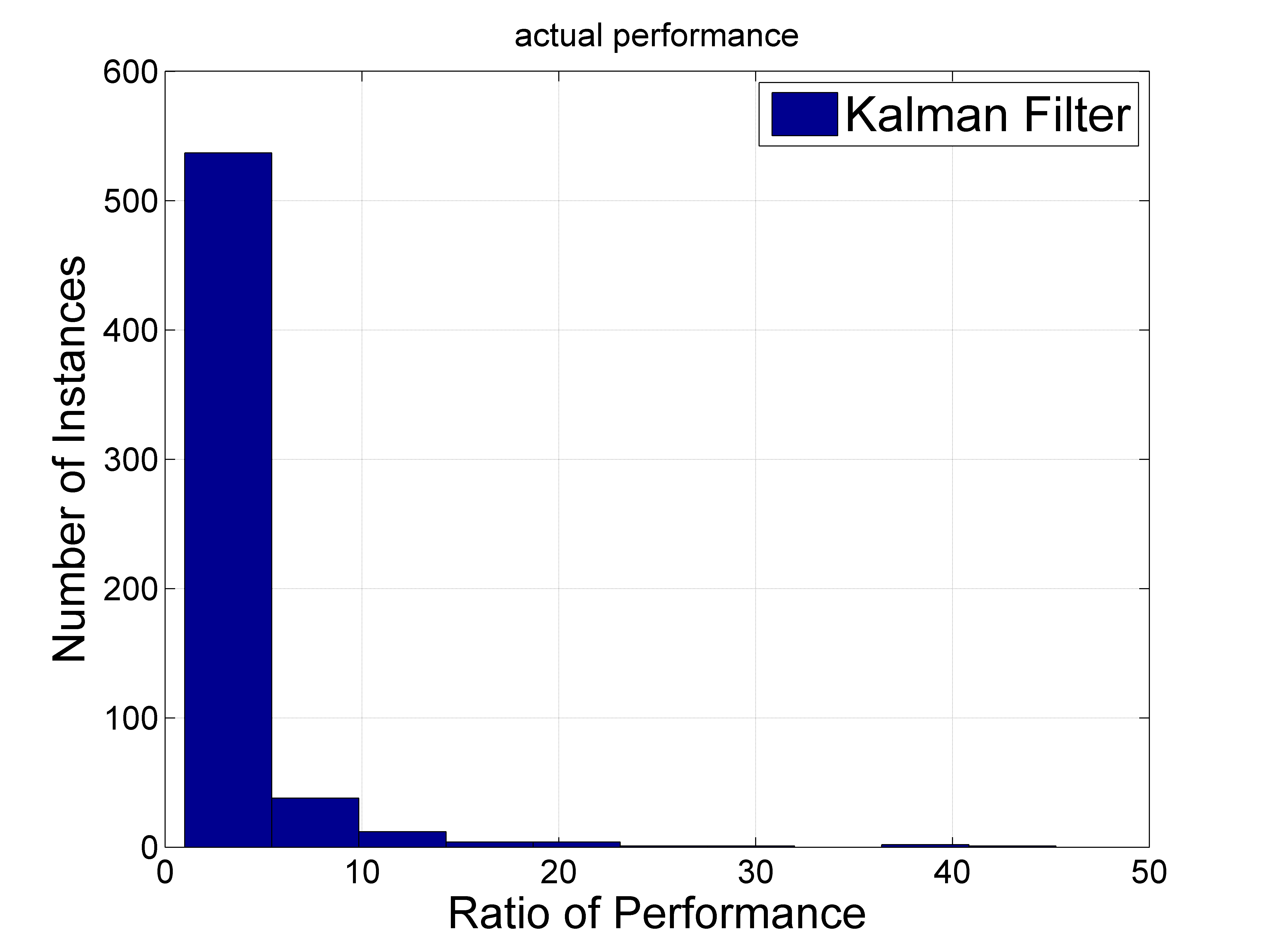}
 \caption{Performance results for greedy algorithm when applied to KFSS problem with $|Q|=10$}
 \label{f:600_10_4}
\end{figure}

\begin{figure}[htb]
 \includegraphics[width=90mm]{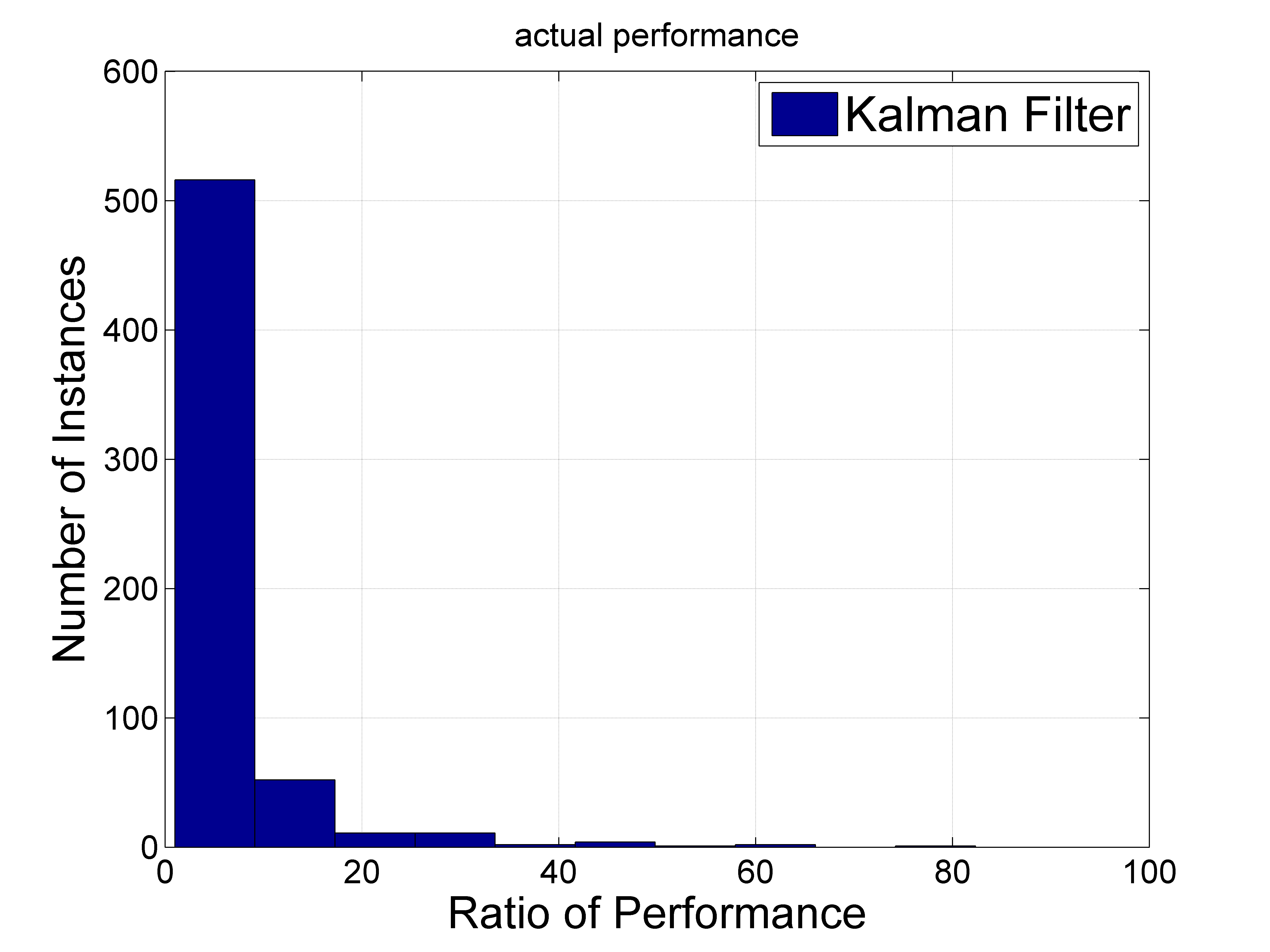}
 \caption{Performance results for greedy algorithm when applied to KFSS problem with $|Q|=20$}
 \label{f:600_20_4}
\end{figure}

\begin{figure}[htb]
 \includegraphics[width=90mm]{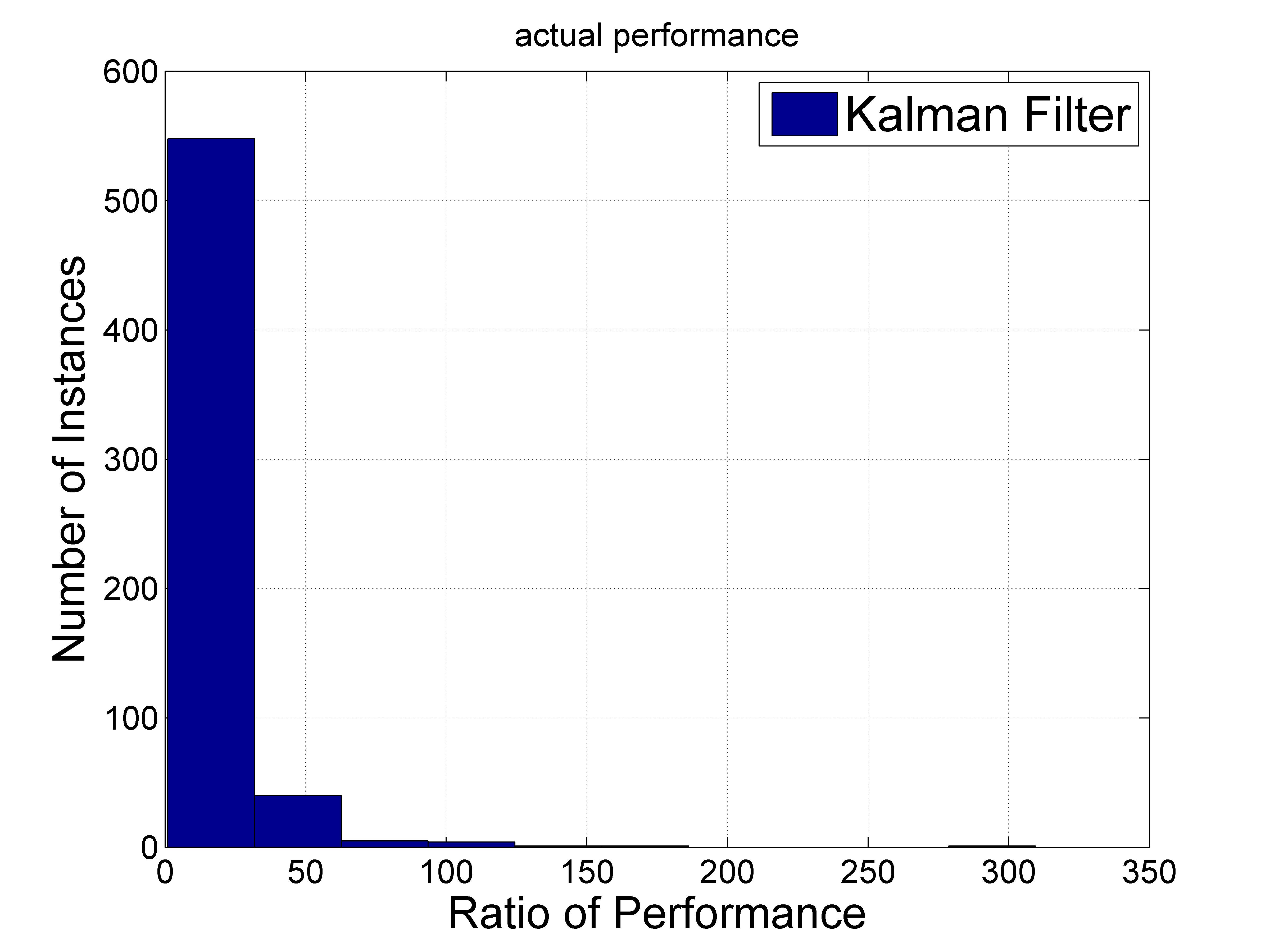}
 \caption{Performance results for greedy algorithm when applied to KFSS problem with $|Q|=30$}
 \label{f:600_30_4}
\end{figure}

The performance results of greedy algorithm for all three cases are displayed in Figure \ref{f:600_10_4}, Figure \ref{f:600_20_4} and Figure \ref{f:600_30_4}, a brief summary of the result is presented in Table \ref{tab:res_sum}, and numerical interpretation of the result is displayed in Table \ref{tab:res_interpret}. In Table \ref{tab:res_sum} $b(i)$ represents the number of systems out of the total number of system tested that has performance ratio lower than $i$, and $a(10)$ represents the number of systems which performance ratio is higher than 10. In Table \ref{tab:res_interpret}, $\mu_r$ refers to the average performance ratio of greedy algorithm on KFSS problem for the given case, $C_v(r)$ refers to the coefficient of variation of the performance ratio, and max$(r)$ refers to the largest performance ratio among the ones the greedy solutions have .

\begin{table}
\centering
\begin{tabular}{ | m{1.5cm} | m{0.7cm}| m{0.7cm} | m{0.7cm} |m{0.7cm} |m{0.7cm} |m{0.7cm} |} 
\hline
 & $b(2)$ & $b(4)$ & $b(6)$ & $b(8)$ & $b(10)$ & $a(10)$ \\ 
\hline
$|Q|=10$ & 410 & 103 & 34 & 17 & 11 & 25 \\ 
\hline
$|Q|=20$ & 313 & 111 & 59 & 24 & 23 & 70  \\ 
\hline
$|Q|=30$ & 272 & 115 & 55 & 31 & 21 & 106 \\
\hline
\end{tabular}
\caption{Summary of performance results} 
\label{tab:res_sum}
\end{table}

\begin{table}
\centering
\begin{tabular}{ | m{1.5cm} | m{1cm}| m{1cm} | m{1cm} |} 
\hline
 & $\mu_r$ & $C_v(r) $ & max$(r)$ \\ 
\hline
$|Q|=10$ & 2.722 & 1.5650 & 45.26 \\ 
\hline
$|Q|=20$ & 4.990& 1.7075 & 82.34 \\ 
\hline
$|Q|=30$ & 9.267 & 2.3213 & 309.5 \\
\hline
\end{tabular}
\caption{Interpretation of performance results} 
\label{tab:res_interpret}
\end{table}

As illustrated in Table \ref{tab:res_sum} and \ref{tab:res_interpret}, as $|Q|$ decreases, the number of solutions of greedy algorithms that performs within performance ratio 2 increases, and the number of greedy algorithm solutions that performs worse than performance ratio 10 decreases. We can also observe from Table \ref{tab:res_sum} that as $|Q|$ decreases, the coefficient of variation and the maximum performance ratio decreases. In other words, as $|Q|$ decreases, not only does the performance of greedy algorithm becomes better in general, the quality of its solutions also becomes less varied and the worst case performance ratio also decreases.

\section{Conclusion and Future Work}
In this work, we studied the performance of greedy algorithm when applied to KFSS problems. Using the fundamental characteristics of greedy algorithm, we provided an improved upper bound for the worst case performance of greedy algorithm when the systems are stable and and have totally ordered error covariance matrices and uncorrelated sensor noise, and highlighted the system parameters that affects its performance. Then, we provided a category of system parameters where greedy algorithm always produces optimal results. For further studies on determining how system parameters affects the performance of greedy algorithm for KFSS problems with diagonally dominant sensor noise covariance matrix is of interest.

\addtolength{\textheight}{-12cm}   




\section*{ACKNOWLEDGMENT}

The author thank Professor Shreyas Sundaram, Professor Dengfeng Sun, Mr. Dawei Sun and Mr. Zhe Wang for their useful comments in this topic.

\begin{minipage}{\linewidth}

\begin{appendix}

\subsection{Miller's Theorem of Matrix Inversion}

For nonsingular matrices A and B where rank of B, $r$, is larger than 1, We can rewrite B as the summation of $r$ rank 1 matrices $(B=\sum_{i=1}^r B_i)$. Then, if we can set $C_1=A$ and every $C_{k+1}=A+\sum_{i=1}^k B_i$ is nonsingular, we can write the inverse of $C_{k+1}$ as 

\begin{equation}
\label{Miller1}
C_{k+1}^{-1}=C_{k}^{-1}-g_k C_{k}^{-1} B_{k+1} C_{k}^{-1}
\end{equation}

Where $g_k=\frac{1}{1+trace(C_{k}^{-1} B_{k+1})}$

Then the inverse of $(A+B)$ can be expressed as

\begin{equation}
\label{Miller2}
(A+B)^{-1}=C_{r}^{-1}-g_r C_{r}^{-1} B_r C_{r}^{-1}
\end{equation}

\subsection{Matrix Inequality Lemma 1}
For any $A_{n \times n}$, $B_{n \times n}$ and $C_{n \times n}$, if $B-C$ is positive semi-definite then $trace(A(B-C)A^T) \geq 0$.

Proof: Since $B-C$ is positive semi-definite, $B-C$ can be Cholesky decomposed into the matrix product of a lower triangular matrix $L$ and its transpose, $B-C=L L^T$. Therefore, $trace(A(B-C)A^T)=trace(A L (A L)^T) \geq 0$.

\subsection{Matrix Inequality Lemma 2}

If $A$,$B$,$C$,$D$ are positive semi-definite and both $A-B$ and $C-D$ are positive semi-definite, then $AC-BD$ is positive semi-definite.

Proof: $B(C-D)=BC-BD \succeq 0$, and $(A-B)C=AC-BC \succeq 0$. Therefore, $AC-BC+BC-BD=AC-BD\succeq 0$.

\subsection{Matrix Inequality Lemma 3}

Given two positive definite matrices $A$ and $B$, 

$\lambda_{min}(A)\lambda_{min}(B) \leq \lambda_{min}(AB)$

proof:

$|| (AB)^{-1} ||_2 \leq ||A^{-1}||_2 ||B^{-1}||_2=\lambda_{max}(A^{-1}) \lambda_{max}(B^{-1})$, which leads to $\frac{1}{\lambda_{min} (AB)} \leq \frac{1}{\lambda_{min}(A)} \frac{1}{\lambda_{min}(B)}$, which in turn leads to $\lambda_{min}(A)\lambda_{min}(B) \leq \lambda_{min}(AB)$.

\subsection{Matrix Inequality Lemma 4}

If $A$ and $B$ are both positive semi-definite and $A-B$ is positive semi-definite, $\lambda_{min}(A) \geq \lambda_{min}(B)$.

Proof:
If for some $A$ and $B$ that satisfies $A-B \succeq 0$ we have $\lambda_{min}(A) < \lambda_{min}(B)$, then for the eigenvector $x$ of $A$ that corresponds to $\lambda_{min} (A)$, we have $ x^T A x-x^T B x \leq \lambda_{min}(A) |x|^2 -\lambda_{min} (B) |x|^2 <0$, which violates the condition $A-B \succeq 0$.

\subsection{Matrix Inequality Lemma 5}

If $A$, $B$ are positive definite matrices, and $A-B \succeq 0$, then $B^{-1}-A^{-1} \succeq 0$. \cite{Argerami}

Proof:
Since $A-B \succeq 0$, $B^{-\frac{1}{2}}A B^{-\frac{1}{2}}-I \succeq 0$, which can be rewritten as $B^{-\frac{1}{2}}A^{\frac{1}{2}} A^{\frac{1}{2}} B^{-\frac{1}{2}}-I \succeq 0$. Since eigenvalues of $(B^{-\frac{1}{2}}A^{\frac{1}{2}}) (A^{\frac{1}{2}} B^{-\frac{1}{2}})$ are the same as the eigenvalues of $(A^{\frac{1}{2}}B^{-\frac{1}{2}})  (B^{-\frac{1}{2}}A^{\frac{1}{2}})$, by applying Matrix Inequality Lemma 4 to the above inequality we find that $(A^{\frac{1}{2}}B^{-\frac{1}{2}})  (B^{-\frac{1}{2}}A^{\frac{1}{2}}) \succeq I$, which leads to $B^{-1}-A^{-1} \succeq 0$.

\end{appendix}
\end{minipage}

\end{document}